# Improving Performance of Federated Learning based Medical Image Analysis in Non-IID Settings using Image Augmentation

Alper Emin Cetinkaya
*Information Security Program*
*Gazi University*
Ankara, Turkey
aemin.cetinkaya@gazi.edu.tr
0000-0003-2424-6075

Murat Akin
*Gazi AI Center of Gazi University,*
*Basarsoft Information Systems Inc.*
Ankara, Turkey
muratakin@gazi.edu.tr
0000-0003-0001-1036

Seref Sagiroglu
*Computer Engineering Dept.*
*Gazi AI Center, Gazi University*
Ankara, Turkey
ss@gazi.edu.tr
0000-0003-0805-5818

*Abstract*—Federated Learning (FL) is a suitable solution for making use of sensitive data belonging to patients, people, companies, or industries that are obligatory to work under rigid privacy constraints. FL mainly or partially supports data privacy and security issues and provides an alternative to model problems facilitating multiple edge devices or organizations to contribute a training of a global model using a number of local data without having them. Non-IID data of FL caused from its distributed nature presents a significant performance degradation and stabilization skews. This paper introduces a novel method dynamically balancing the data distributions of clients by augmenting images to address the non-IID data problem of FL. The introduced method remarkably stabilizes the model training and improves the model's test accuracy from 83.22% to 89.43% for multi-chest diseases detection of chest X-ray images in highly non-IID FL setting. The results of IID, non-IID and non-IID with proposed method federated trainings demonstrated that the proposed method might help to encourage organizations or researchers in developing better systems to get values from data with respect to data privacy not only for healthcare but also other fields.

## I. INTRODUCTION

Using deep learning (DL) for medical image analysis such as detecting Covid-19 from the chest X-Ray (CXR) images without using any dedicated test kits is a low-cost and accurate alternative to laboratory-based testing. Recent advances in DL provides promising results for medical image analysis and large-scale diagnostic. Also, the ease of access, scalability, and the rapid diagnostic of DL are huge plus when compared to human based diagnosis. However, DL methods require large amounts of samples to achieve competitive results since the performance of the DL algorithms is highly affected by the volume and diversity of the data.

The approach of centrally training an DL model for leveraging health data suffers from the risk of violation of patient privacy with the increasing concerns on data privacy. Also, it is usually not likely that medical institutions share their local data due to ownership concerns and strict regulations on data privacy such as Health Insurance Portability and Accountability Act (HIPAA) and General Data Protection Regulation (GDPR). Even if the collected data is adequately protected against malicious actors, it is high probability that a condition beyond expected situations resulting the violation of individuals privacy may occur. The healthcare data including personal identity, behavior, biometrics, biomedical images, genomic data, and medical history of patient has become the one of the primary targets of attackers or hackers while the healthcare is the sector that is most exposed to cyber-attacks. According to the recent report published by HIPAA [1], healthcare records of more than 5 million people were breached across 38 incidents in August 2021, while these leaks bring the total figure to 707 in the period between September 2020 and August 2021. The breach of the health data has a lifelong impact unlike any other personal data breach since it may include information such as genomic data that cannot be altered afterwards. Data holders, who are obliged to ensure the security and privacy of the data they keep, face serious economic and legal consequences in such cases. Hence, one of the primary challenges on developing data-driven intelligent applications for healthcare is to preserve privacy and secure shared data against any kind of cyber threats or attacks.

A naive solution to leverage high volume and diversity of data across multiple organizations is to alter the data before collecting to a central place eighter by removing or anonymizing personal data such that no private information of individuals can be inferred. Unfortunately, re-identification of anonymized or removed information against such protections is still possible using advanced attacks [2] such as linkage attacks. Furthermore, there is a trade-off between the data utility and the privacy for this kind of privacy-preserving methods such that the utility of the data decreases as the more privacy is needed [3].

An alternative solution to data anonymization is training a global model with a recent approach called Federated Learning (FL) that was introduced by Google in 2016 [4]. In contrast to conventional strategy of training a model centrally, FL enables collaborative training of a global model across multiple agents without gathering the data to a central place. Instead of gathering the data in a central place, the model training phase is decentralized and the training is performed on the device where the data is produced. Since the data never leaves its origin, there is no need to concern about privacy risks and legal regulations to leverage the high volume and diverse data. FL also reduces the cost of hardware that is needed to store, secure, and process all the mass data gathered from numerous devices.

FL differs from traditional distributed learning where user data is aggregated on a centralized location and then distributed to the agents in an independent and identical distribution [5],[6]. Unlike centralized or traditional distributed learning, assumption of IID data practically does not for real-world scenarios. Generally, each client device generates data by a distinct distribution and the number of data being generated is also highly unbalanced. There are basically three types of non-IIDness that are given below with example cases [7]:

i. Label distribution skew: Hospitals having specialized in specific diseases have more and diverse data on them,
ii. Feature distribution skew: Same disease showing different symptoms and effects in patients,
iii. Quantity distribution skew: Variable number of local medical images varies across hospitals.

Beside from the cases mentioned above, same labels for different features or same features for different labels also presents a heterogeneity in data distribution. It has been observed from [8] and [9] that the performance of FL is deeply dependent on the local data distribution of the clients. Hence, the great degree of statistical heterogeneity of data distribution presents significant degradations to the accuracy and the convergence.

In this study, we proposed a method to dynamically balance the local data distributions of clients using augmenting extra images to mitigate the performance and stabilization degradations caused by non-IID data. Through experiments, we compared the results of IID, non-IID and non-IID data with proposed method for FL.

The main contributions of this work are summarized as:

- Proposing a novel method to alleviate performance degradation caused by the non-identical and independent data distribution in FL.
- Showing experimentally that non-IID data presents significant challenges in stability and the accuracy of FL and presenting experimental comparison for the performance of FL in IID, non-IID and the non-IID with proposed method settings.
- Implementing successfully centralized medical data with FL on highly non-IID setting and supporting privacy preserving medical image analysis.
- Sharing publicly the source code of our proposed method [10].

## II. BACKGROUND

In this section, we briefly introduce FL and core challenges arise from the distributed nature of it.

### A. Federated Learning (FL)

FL is a method by which many client devices train a global model collaboratively using their local data under the control of a central server [4]. The ability to train a global model without gathering all the mass data from the vast number of clients makes FL a suitable setup for data privacy applications without concerning privacy risks and legal regulations.

The main objective in FL is to minimize the cost function [4] given below across a distributed network of clients. $K$ is the total number of clients, $p_k$ is relative impact of device $k$ where $\sum_{k=1}^{K} p_k = 1$ and the $f_k(\omega)$ is the local objective function of $k^{\text{th}}$ client.

$$\min_{\omega} F(\omega), \text{ where } F(\omega) \coloneqq \sum_{k=1}^{K} p_k f_k(\omega) \quad (1)$$

A typical round of a FL method [4] is given in Fig. 1. The steps are repeated until the desired success rate is achieved in the global model.

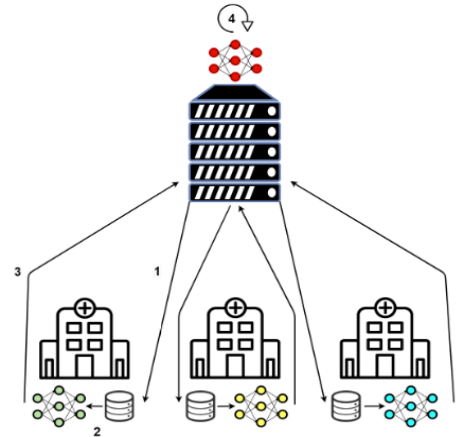

Fig. 1. Overview of FL framework (borrowed from [11]) (1) Central server selects a random set of clients and global model is sycnhronized with them, (2) Clients perform local training using their local data, (3) Clients upload their local updates to the server, (4) Updates from the clients are aggregated by the server to improve the global model

FederatedAveraging (FedAvg) [12] is the method for model aggregation in which the server takes the weighted average of the trained models to update the global model. In each round of FedAvg, the selected clients perform local computations in parallel for the $E$ epochs to obtain the updated weights $\omega_{t+1}^k$ for client update phase. After collecting the updated weights from the clients, the server performs the FedAvg operation given in Eq. 2 to improve the global model.

$$\omega_{t+1}^k \leftarrow \text{ClientUpdate}(\omega_t, k), and \ \ \omega_{t+1} \sum_{k=1}^{K} \frac{n_k}{n} \omega_{t+1}^k \quad (2)$$

Averaging the local updates relative to their local sample counts provides resistance against performance degradation of imbalanced local data. In our work, we also examine the performance of FedAvg algorithm in non-IID data and compared the results with the simple averaging without considering relative effects of local updates.

### B. Core Challenges

The core challenges [13],[14] associated with the distributed nature of FL are described below. It should be emphasized that we mainly focus on the statistical heterogeneity and privacy in FL so the other challenges are out of scope of this work.

**Statistical Heterogeneity**: The local data distribution of the clients varies greatly among each other and the individual distributions is usually different from the overall distribution. The non-IIDness that is present in local data distributions of clients causes significant performance issues for FL.

**Privacy**: Although the data is kept locally and not shared with the server, information about the client's data can be inferred from updates using advanced attacks [15]. Privacy for FL can be provided with the integration of other privacy preserving techniques such as secure multi-party computation [16] or differential privacy [17].

**Communication Cost**: Since the number of clients contributing the training of a global model is massive and the global model and updates are exchange between clients and central server multiple times, the communication cost of downloading from and uploading to server is usually the primary bottleneck in FL both for clients and the server.

**Systems Heterogeneity**: Similar to statistical heterogeneity, the computational, storage or communicational resources of clients varies significantly across network which may cause the stragglers in rounds also presents a challenge to tackle for improved FL.

## III. RELATED WORK

In FL, the statistical heterogeneity among the clients is one of the primary challenges as mentioned previously. The local objective of each client device differs from the global objective since each client has a local dataset highly different from each other resulting a drift in the local updates [18]. Therefore, the non-IID data causes the aggregated model to be far from the global optimum presenting the performance degradation when compared with the IID assumption. To improve the performance of FL in non-IID setting, there are various studies [18-21] focusing on different methods to alleviate the skews caused from the non-IID data.

Zhao et. al. [18] investigated the accuracy drop of FL originated from the non-IID data and showed that this reduction can be explained by the weight divergence as shown in Fig. 2. To cope with the statistical heterogeneity challenge, they proposed a method in which the central server shares a global proxy dataset containing uniform distribution over labels with all clients to balance the local data distributions and they also showed through experiments that their proposition significantly increases the test accuracy. Also, they proposed distributing a warm-up model trained on the global dataset to clients instead of randomly initialized weights.

Lie et. al. [19] introduced a framework, inspired by the FedAvg, named FedProx to address the challenges of heterogenous data distribution in FL without any computational overhead and privacy concession. They made modifications to the FedAvg and added a $L_2$ regularization term to the local loss functions to limit the distance between the local updates and the global model. As a result of proposed improvements, FedProx demonstrates significantly more stable and robust convergence behavior compared to FedAvg in highly heterogenous settings. They also addressed the systems heterogeneity in FL by allowing each client to perform variable epochs of local training based on their system resources.

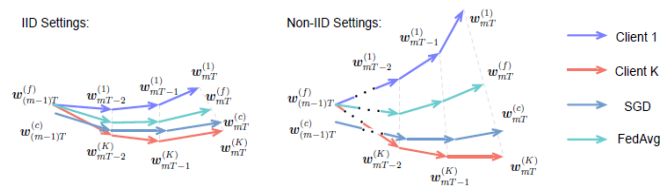

Fig. 2. Illustration of the weight divergence for FL with IID and non-IID data (borrowed from [18]). SGD refers to the centralized setting.

Wang et. al. [20] proposed an algorithm called FedNova (Federated Normalized Averaging Algorithm) that also allows clients to perform variable local training iterations in each round. The clients that perform a larger number of local trainings will have a greater effect on the global model when they are simply averaged with other updates in the round. Thus, to reduce the bias originated from the variable number of local trainings and non-IID data of the clients, FedNova scales and normalizes the local gradients before aggregation instead of simply averaging them to update the global model.

Yao et. al. [21] proposed a method called FedMMD (Maximum Mean Discrepancy) to alleviate the performance degradation caused by the non-IID local data in FL. FedMMD is a two-stream model that consists of a locally trained model and a global model and presents an MMD constraint between them so that the local model is forced to integrate further knowledge from the global model in the training phase.

## IV. MATERIAL AND METHOD

In this section, we depict the details of the dataset, our method for artificially distributing the dataset in a non-IID way to multiple clients, reference model architecture for chest diseases classification and the proposed method for mitigating the performance degradation caused from non-IID distribution by balancing the local data distributions dynamically using image augmentation.

### A. Dataset Description

Here, we explain the aggregated dataset used in our work which is composed of 28833 CXR images with four different labels from 5 different publicly available datasets [11]. Our dataset contains a mixture of Covid-19, non-covid viral or bacterial cases of pneumonia, normal and lung opacity cases. Number of samples for each label in the aggregated dataset used in our work are given in Table 1. Also, Fig. 3 represents one of the examples borrowed from the datasets. The arrow in the figure depicts the risky or anomaly region at the data. Since we shared the total number of samples between the clients to simulate the federated setting, the number of samples should be quite enough for a realistic real-world scenario.

TABLE I. DETAILED NUMBERS FOR THE DATASET

| Labels | Covid-19 | Pneumonia | Lung Opacity | Normal |
|---|---|---|---|---|
| # of Instances | 5429 | 5618 | 6011 | 11775 |

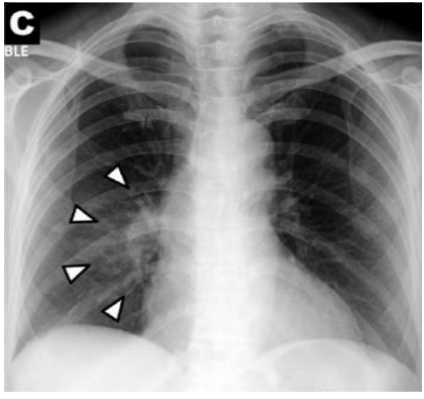

Fig. 3. A sample chest X-ray image for a person infected by Covid-19 (Borrowed from [22])

## B. Reference Model Architecture

Due to its promising performance and various advantages such as fast diagnosis, easy scalability and accessibility, the CNN architectures have been extensively used in aiding medical image analysis.

As a reference model, we use a CNN architecture that we developed for the classification of the multiple diseases from CXR images [11]. CNNs are one of the most popular classes in neural networks and used in dealing with images and visual recognition tasks. A CNN is usually a combination of convolutional, pooling, and fully connected layers [23] based on mathematics. The block diagram of the reference CNN model is shown in Fig. 4. As clearly seen from the figure, the model consists of 5 convolutional layers followed by 2 fully connected layers. The kernel size of the first layer is 7x7, 5x5 for the second layer, and 3x3 kernels for the other convolutional layers. Furthermore, the 2x2 Max pooling layer and dropout with 0.1 probability to avoid overfitting and batch normalization layers are added after each convolutional layer.

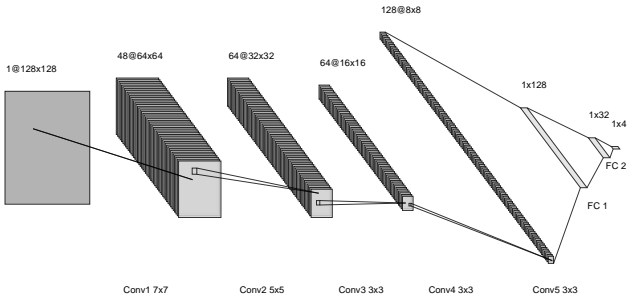

Fig. 4. Reference CNN architecture for multi-disease classification from CXR images. (Each convolutional layer is follewed by 2x2 Max Pooling, Batch Normalization and Dropout layers)

## C. Artificial Non-IID Partitioning

We use Dirichlet distribution to artificially distribute the central dataset in a non-IID fashion to each client for FL simulation over statistically heterogeneous networks.

The probability density function of Dirichlet distribution with parameters $\alpha_1, \dots, \alpha_k > 0$ is given in (3). The $\alpha$ parameter of the distribution is used to control the imbalance level of the values sampled from the distribution. As can be seen from the equation, the imbalance level of sampled values is increases as the corresponding $\alpha$ parameters increase.

$$f(x_1, \dots, x_k; \alpha_1, \dots, \alpha_k) = \frac{1}{B(\alpha)} \prod_{i=1}^{k} x_i^{(\alpha_i - 1)} \qquad (3)$$

where $B(\alpha) = \frac{\prod_{i=1}^{k} \Gamma(\alpha_i)}{\Gamma(\sum_{i=1}^{k} \alpha_i)}$ and $\sum_{i=1}^{k} x_i = 1$

For each label in the central dataset, we sampled the 20 variable Dirichlet distribution with $\alpha_1, \dots, \alpha_{20} = 1$ parameters corresponding to the 20 individual hospitals and distributed the sampled amount of data for corresponding label to clients to artificially partition the central dataset in a non-IID way. The resulting local data distribution of clients is shown in Fig. 5. As can be seen from the figure, there is a high feature and quantity distribution skews are present among the clients.

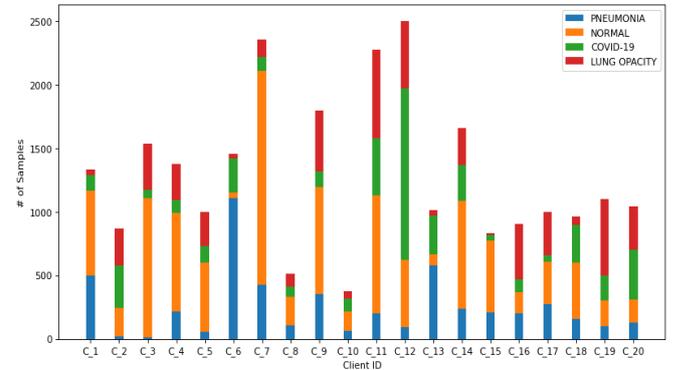

Fig. 5. Local data distributions for a non-IID setting

## D. Proposed Mitigation Method

To alleviate the skew caused by the non-IID distribution, we propose a method that balances the distribution of the selected clients' dataset for each round dynamically. We added an augmentation phase to the vanilla FedAvg algorithm. Image augmentation is a process of artificially creating training images through various transformation such as randomly rotating the image or changing the brightness values to enhance the available number of samples for training.

In each round of FL, the selected clients send the number of samples for each label on their local set. Afterwards, server selects the number of maximum samples among clients for each label and send this information to each client selected for the round. Then, the clients generate synthetic images by applying 14 different transformations to their local dataset and increase their number of samples for each label to the required number. If the required number of samples is higher than the number that can be reached by applying these transformations individually, the combination of them is applied until the desired number of samples is achieved. The complete transformations and their effects on original image are shown in Fig. 6. After the augmentation phase, each client in round will have a same local data distribution with equal number of samples for each label.

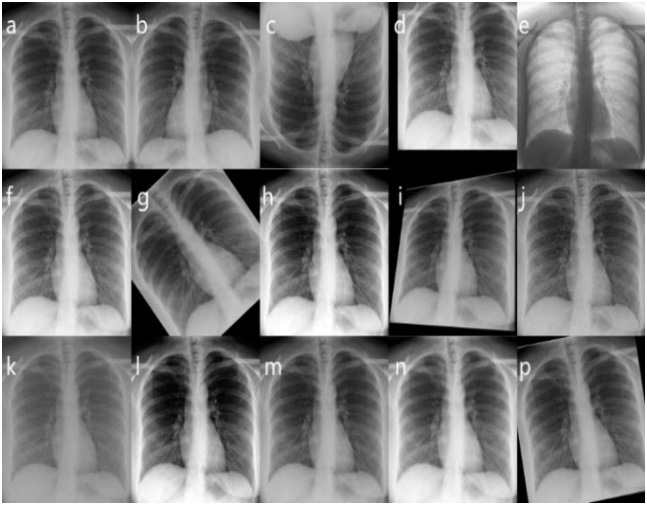

Fig. 6. Transformations applied for generating syntethic images. (a) Original, (b) Horizontal flip, (c) Vertical flip, (d) Crop, (e) Invert, (f) Solarize, (g) Rotation, (h) Jitter, (i) Perspective, (j) Sharpness, (k) Gaussian noise, (l) Histogram equalization, (m) Contrast, (n) Gaussian blur, (p) Affine.

## V. EXPERIMENTAL STUDIES

As explained in Section 4 in the details of the proposed approach, the reference model architecture, DL Model, and non-IID partitioning, the experimental details are given here. The settings and the results of FL models achieved with different setups are presented and achieved results are demonstrated.

### A. Setting

To evaluation purposes, we split the central dataset mentioned above into train and test sets of 90% and 10%, respectively. The test set consisting of 2884 CXR images mixture of 574 Covid-19, 536 non-covid viral or bacterial pneumonia, 620 lung opacity, 1154 and healthy cases. Also, we used the same training and test sets for each experiment.

To simulate the FL environment, we distributed the central training set in different ways to $K = 20$ clients corresponding the data owner medical organizations. For federated local trainings, all clients perform a single epoch of local work, $E = 1$, with a batch size of $B = 16$. In each round of FL, 5 clients are randomly selected. The reference CNN model is used for all trainings and the weight optimization method used in the client side to update the weights is Adadelta optimization with a learning rate of 0.005. For the model aggregation phase, we use FedAvg approach given in Eq. 2. We measure the test accuracy of global model on centrally located test set after each round of FL.

### B. Results

For tests, firstly, we partitioned the central training dataset with IID way so that each client had equal number of samples from the uniformly shuffled local training set. The FL with IID setting achieved the classification task on test set with 89.22% accuracy after 100 communication rounds. Then we distributed the central training set in a non-IID fashion using the artificial non-IID sharing with Dirichlet distribution for $\alpha_1, \ldots, \alpha_{20} = 1$ parameters and trained a global model for 100 rounds. The developed FL model with non-IID data achieved the task with 83.22% test accuracy having a significant drop. Lastly, we added our proposed mitigation method in FL and again trained a global model for 100 communication rounds. The global model obtained after federated training over non-IID data with the proposed method achieved the task with 89.43% accuracy. The proposed method even outperformed with the IID setting since the number of training samples was also increased at the same time when balancing the local distributions. The test results for these three methods are shown in Fig. 7. As can be seen from the figure, the non-IID data brought significant increase in performance issues based on FL approach and our proposed method alleviated the degradation caused by non-IID data. The final accuracy of FL approach was also improved.

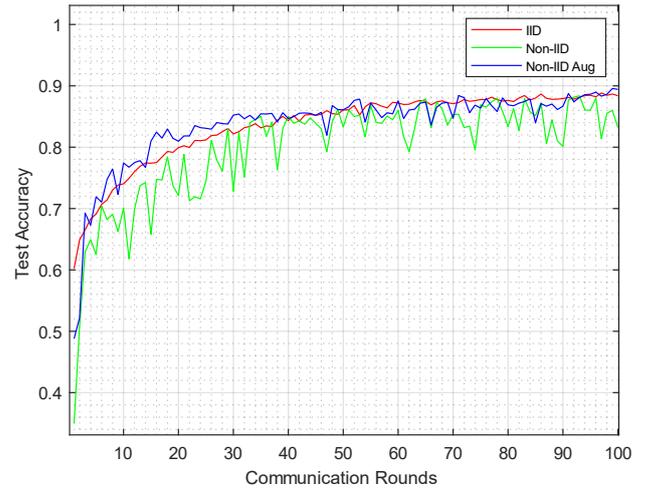

Fig. 7. Comparisons for test results with 100 rounds of FL models with high non-IID setting

Secondly, we increased all the $\alpha$ parameters of the Dirichlet distribution to 2 and repeated the same experiment to evaluate the performance of the proposed method with less imbalanced non-IID data that was a much closer to non-IID case. Resulting local data distributions of the clients is shown in Fig. 8. The performance degradation is less significant in non-IID setting as shown in Fig. 9. However, our proposed method improved the training accuracy from 88.90% to 89.66% and the stabilization improvement was still remarkable.

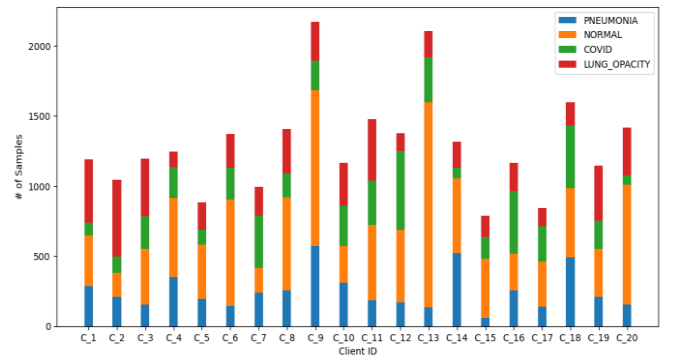

Fig. 8. Local data distributions for less non-IID setting

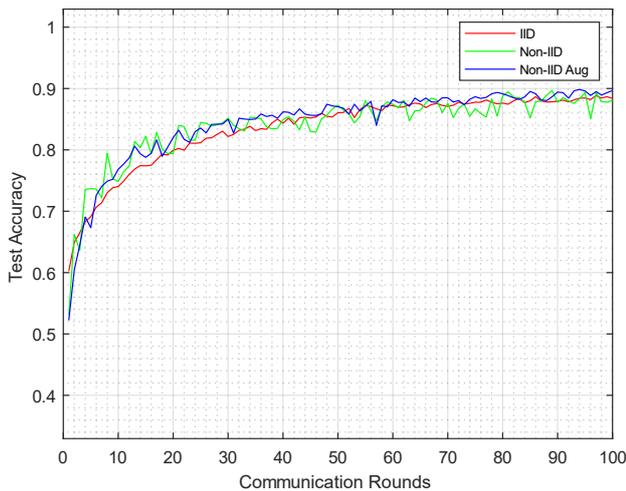

Fig. 9. Comparisons for test results with 100 rounds of FL models with less non-IID setting

## VI. Conclusion And Future Work

In this paper, a novel method dynamically balancing the data distributions of clients by augmenting images to address the non-IID data problem of FL was successfully introduced and improved the ultimate model's accuracy from 83.22% to 89.43% for multi-chest diseases. The results of IID, non-IID and non-IID with the introduced method were demonstrated that the proposed method might help to develop better solutions for not only for healthcare but also other fields.

Data utility and privacy are another important trade-off to be considered to improve model accuracies or architectures of models. Increased performance of FL with non-IID data might encourage the medical institutions to start or adopt their studies with FL approach for developing data driven intelligent solutions to reduce the risk of facing data breaches or fines of privacy regulation.

FL adoption might play a key role for developing DL applications for industries where the concerns about data privacy are still there where large and diverse private data are hard to access and process due to legal regulations. However, the highly unbalanced data of massively distributed clients in real world present a significant challenge for FL.

As stated in [15], keeping data locally and not sharing with the server confronted advanced attacks. Increasing data utility as suggested in this article might help to build up and train DL models in client side using methods such as differential privacy algorithms randomly perturbing data, models or outputs and supporting privacy & utility problem.

The authors will focus on the advanced strategies for balancing the non-IID distribution of FL architecture in the future works. Instead of balancing the local distributions, smart algorithms for client selection might be developed to select the clients having similar data distributions.